\documentclass{iau}
\usepackage{graphicx}
\usepackage{amsmath}
\usepackage{amssymb}
\usepackage{gensymb}
\usepackage[T1]{fontenc} % if needed
\usepackage{dsfont}

\title[Non-Gaussianity \& posterior of inflationary parameters] %% give here short title %%
{Generic inference of inflation models\\ by local non-Gaussianity}

\author[S.\ Dorn, E.\ Ramirez, K.\ E.\ Kunze, S.\ Hofmann, T.\ A.\ En{\ss}lin]   %% give here short author list %%
{Sebastian Dorn$^1$, Erandy Ramirez$^2$, Kerstin E.\ Kunze$^3$, \\Stefan Hofmann$^4$, \and Torsten A.\ En{\ss}lin$^{1,5}$}

\affiliation{$^1$Max-Planck-Institut f\"ur Astrophysik,\\ Karl-Schwarzschild-Str.~1, D-85748 Garching, Germany \\email: {\tt sdorn@mpa-garching.mpg.de} \\[\affilskip]
$^2$Instituto de Ciencias Nucleares, UNAM A. Postal 70-543, Mexico D.F. 04510, Mexico \\[\affilskip]
$^3$Departamento de F\'\i sica Fundamental and IUFFyM, Universidad de Salamanca,\\ Plaza de la Merced s/n, 37008 Salamanca, Spain \\[\affilskip]
$^4$Arnold Sommerfeld Center for Theoretical Physics, Ludwigs-Maximilians-Universit\"at M\"unchen, Theresienstra{\ss}e 37, D-80333 Munich, Germany \\[\affilskip]
$^5$Ludwigs-Maximilians-Universit\"at M\"unchen,\\ Geschwister-Scholl-Platz 1, D-80539 Munich, Germany
}

\pubyear{2014}
\volume{306}  %% insert here IAU Symposium No.
\pagerange{119--126}
\jname{Statistical Challenges in 21st Century Cosmology}
\editors{A.C. Editor, B.D. Editor \& C.E. Editor, eds.}
\begin{document}

\maketitle

\begin{abstract}
The presence of multiple fields during inflation might seed a detectable amount of non-Gaussianity in the curvature perturbations, which in turn becomes observable in present data sets like the cosmic microwave background (CMB) or the large scale structure (LSS). Within this proceeding we present a fully analytic method to infer inflationary parameters from observations by exploiting higher-order statistics of the curvature perturbations. To keep this analyticity, and thereby to dispense with numerically expensive sampling techniques, a saddle-point approximation is introduced whose precision has been validated for a numerical toy example. Applied to real data, this approach might enable to discriminate among the still viable models of inflation.
\end{abstract}

\firstsection % if your document starts with a section,
\section{Motivation \& data model}
Precision measurements of the CMB and the LSS have opened a window to the physics of the early universe (\cite[WMAP 2012]{WM12}, \cite[Planck 2013]{Pl13}). In particular it has become possible to measure the exact statistics of the primordial curvature perturbations on uniform density hypersurfaces, $\zeta$, which turned out to be almost Gaussian (\cite[Planck 2013(2)]{Pl13(2)}). These small deviations from Gaussianity are commonly represented by the parameters $f_\text{NL}$ and $g_\text{NL}$ and allow to write
\begin{equation}
\label{deffnl}
\zeta = \zeta_1 + \frac{3}{5}f_\text{NL}\zeta_1^2 +\frac{9}{25}g_\text{NL}\zeta_1^3 +\mathcal{O}(\zeta_1^4),
\end{equation}
where $\zeta_1$ denotes the Gaussian curvature perturbations.

\begin{figure}[ht]
\includegraphics[width=\textwidth]{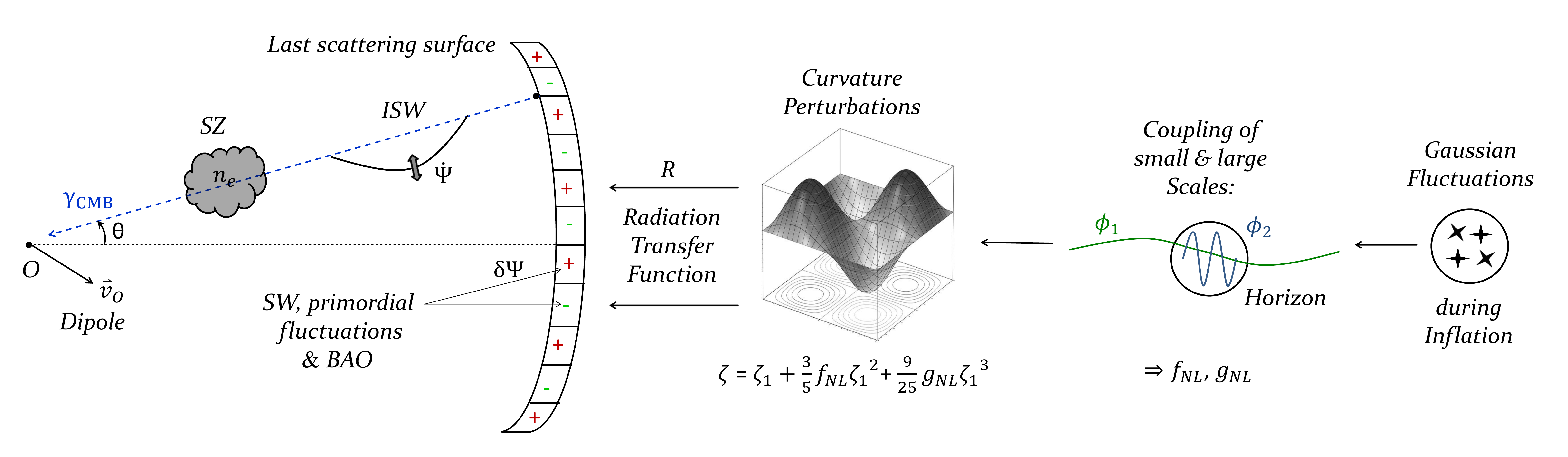}
\caption[width=\textwidth]{Schematic illustration of the generation of non-Gaussian curvature perturbations out of inflation and their translation to the observable CMB.\label{cosmo}}
\end{figure}

On the other side, the exact statistics of $\zeta$ are predicted by inflation models. This suggests to relate current observations directly to inflation models, parametrized by inflationary parameters, $p$, by higher order statistics of the curvature perturbations (see Fig.\ \ref{cosmo}). How to set up such an inference approach in the framework of information field theory (\cite[En{\ss}lin et al. 2011]{En11}) was originally developed in (\cite[Dorn et al. 2013]{Do13}, \cite[Dorn et al.\ 2014]{Do14}) and is addressed in the work at hand.

To do inference of inflationary parameter we consider the observation (CMB, LSS) to be a discrete data set, $d=(d_1,\dots,d_m)^T\in \mathds{R}^m,~m\in\mathds{N}$, given by
\begin{equation}
\label{data}
d=\frac{\delta T}{T_\text{CMB}} = R \zeta + n = R\left(\zeta_1 + \frac{3}{5}f_\text{NL}\zeta_1^2 +\frac{9}{25}g_\text{NL}\zeta_1^3 +\mathcal{O}\left(\zeta_1^4\right)\right) +n,
\end{equation}
where $n$ denotes some Gaussian noise, $P(n)=\mathcal{G}(n,N)\equiv |2\pi N|^{-1/2}\exp(-n^\dag N^{-1}n /2)$, and $R$ the so-called response operator. If we consider the CMB, the latter is a linear operator that transfers the curvature perturbations into temperature anisotropies, i.e. the radiation transfer function, and contains all measurement and instrumental effects.

%======================================================================================================================
\section{Posterior of inflationary parameters}
Since the non-Gaussianity parameters depend on a particular inflation model, the calculation of their posterior,
\begin{equation}
\label{inte}
\mathcal{P}(f_\text{NL},g_\text{NL}|d)\propto \mathcal{P}(f_\text{NL},g_\text{NL})\int\mathcal{D}\zeta_1 \exp\left[-\mathcal{H}(\zeta_1,d|f_\text{NL},g_\text{NL})\right],
\end{equation}
is done first (\cite[Dorn et al.\ 2014]{Do14}). $\mathcal{H}$ denotes the information Hamiltonian, $\mathcal{H(~.~)}\equiv -\ln \mathcal{P}(~.~)$. A straightforward calculation of this Hamiltonian shows that it contains terms up to $\mathcal{O}(\zeta_1^4)$. Therefore the integral of Eq.\ (\ref{inte}) cannot be performed analytically. However, this obstacle can be circumvented by conducting a saddle-point approximation in $\zeta_1$ around $\bar{\zeta_1}\equiv \arg\min\left[\mathcal{H}(\zeta_1,d|f_\text{NL},g_\text{NL})\right]$ up to the second order in $\zeta_1$ to be still able to perform the path-integration analytically. This means, we replace $\mathcal{P}(\zeta_1,d|f_\text{NL},g_\text{NL})$ by the Gaussian $\mathcal{G}(\zeta_1 - \bar{\zeta_1}, D_{d,f_\text{NL},g_\text{NL}})$, with
\begin{equation}
D^{-1}_{d,f_\text{NL},g_\text{NL}}\equiv ~\frac{\delta^2 \mathcal{H}(\zeta_1,d|f_\text{NL},g_\text{NL})}{\delta \zeta_1^2}\bigg\vert_{\zeta_1=\bar{\zeta_1}},~0=~\frac{\delta \mathcal{H}(\zeta_1,d|f_\text{NL},g_\text{NL})}{\delta \zeta_1}\bigg\vert_{\zeta_1=\bar{\zeta_1}}. 
\end{equation}
Including the saddle-point approximation and performing the path integral in Eq.\ (\ref{inte}) yields the final expression of the posterior (\cite[Dorn et al.\ 2014]{Do14}),
\begin{equation}
\label{end}
\mathcal{P}(f_\text{NL},g_\text{NL}|d)\propto \sqrt{|2\pi D_{d,f_\text{NL},g_\text{NL}}|} \exp\left[-\mathcal{H}(d,\bar{\zeta_1}|f_\text{NL},g_\text{NL})\right]\mathcal{P}(f_\text{NL},g_\text{NL}).
\end{equation}
Eq.\ (\ref{end}) enables to calculate the posterior of $f_\text{NL},g_\text{NL}$ fully analytic without expensive Monte Carlo sampling techniques. This analyticity has been conserved by conducting a saddle-point approximation, whose sufficiency has been validated by the DIP test (\cite[Dorn et al. 2013(2)]{Do13(2)}). Note that $\mathcal{H}(d,\bar{\zeta_1}|f_\text{NL},g_\text{NL})$ as well as $D_{d,f_\text{NL},g_\text{NL}}$ denpends on the two point correlation function of $\zeta_1$ and thus requires some a priori knowledge on the primordial power spectrum. One might use the currently measured and therefore well motivated primordial power spectrum (pure power law) with best fit parameters from \textit{Planck} (\cite[Planck 2013(3)]{Pl13(3)}).

\begin{figure}[ht]
\includegraphics[width=.5\textwidth]{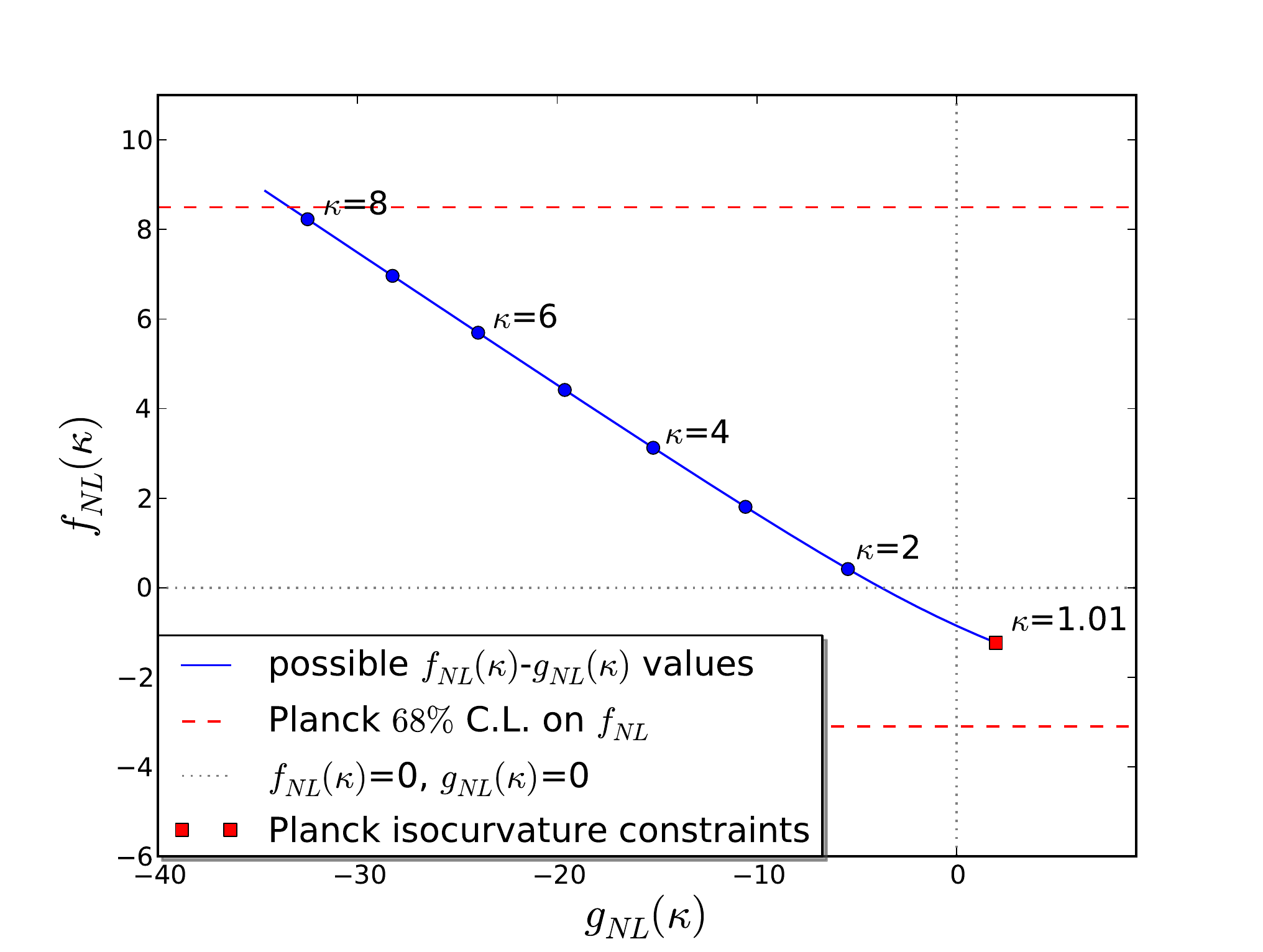}%
\includegraphics[width=.5\textwidth]{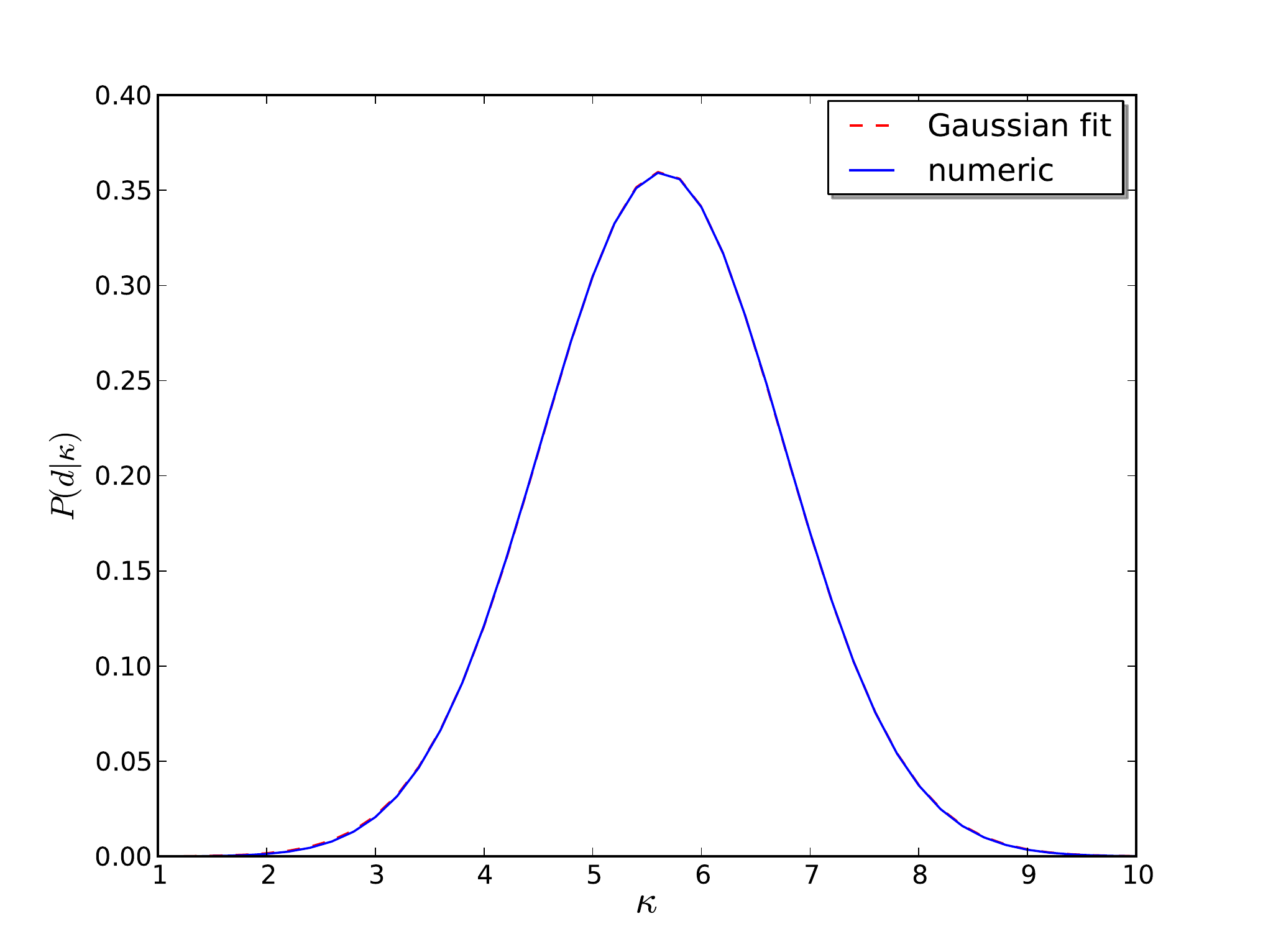}
\caption[width = \textwidth]{Simplest curvaton model: (Left) Possible values of $f_\text{NL}$ and $g_\text{NL}$ within current \textit{Planck} constraints parametrized by the curvaton parameter $\kappa$. (Right) Normalized likelihood distributions for $\kappa$ in a two-dimensional test case in the Sachs-Wolfe limit with data generated from $\kappa_\text{gen}=5$. Figures taken from (\cite[Dorn et al.\ 2014]{Do14}).\label{figgf}}
\end{figure}

To obtain the posterior of inflationary parameters we replace $f_\text{NL}(p),g_\text{NL}(p)$ by their parameter dependent expressions, predicted by inflation models, e.g., for the simplest curvaton model with potential $V(\phi,\chi)=\frac{1}{2}m_\phi \phi^2 + \frac{1}{2}m_\chi \chi^2$~(\cite[Bartolo et al. 2002]{Ba02}, \cite[Sasaki et al. 2006]{Sa06}) one obtains
\begin{equation}
%\begin{split}
f_\text{NL}=\frac{5}{4}\kappa-\frac{5}{3}-\frac{5}{6\kappa},\hspace{.8cm}
g_\text{NL}=\frac{25}{54}\left(-9\kappa +\frac{1}{2} +\frac{10}{\kappa}+\frac{3}{\kappa^2}\right),~
\hspace{.8cm}~\kappa \equiv  \frac{4\bar{\rho}_r}{3\bar{\rho}_\chi} +1. 
%\end{split}
\end{equation}
The posterior for the curvaton parameter $\kappa$ for a toy example in two dimensions as well as possible values of $f_\text{NL}(\kappa),g_\text{NL}(\kappa)$ are illustrated in Fig.\ \ref{figgf}.

%======================================================================================================================
\section{Conclusions}
We presented a novel and generic method to infer inflationary parameters from observations (CMB, LSS) by local non-Gaussianity. The method is fully analytic and thereby avoids expensive sampling techniques. The introduced approximation, necessary to conserve the analyticity, has been validated successfully.

%======================================================================================================================

%======================================================================================================================

\begin{thebibliography}{}
\bibitem[WMAP 2012]{WM12}{WMAP Collaboration} 2013, \textit{APJS}~208~20, arXiv:astro-ph/12125225
\bibitem[Planck 2013]{Pl13}{Planck Collaboration} 2013, arXiv:astro-ph/13035076 %Cosm. Parameter
\bibitem[Planck 2013(2)]{Pl13(2)}{Planck Collaboration} 2013, arXiv:astro-ph/13035084 %NG
\bibitem[En{\ss}lin et al. 2011]{En11}{En{\ss}lin, T.A., Frommert, M., Kitaura, F.S.} 2011, \textit{Phys.~Rev.~D}~80~105005, arXiv:astro-ph/08063474
\bibitem[Dorn et al. 2013]{Do13}{Dorn, S., Oppermann, N., Khatri, R., En{\ss}lin, T.A.} 2013, \textit{Phys.~Rev.~D}~88~103516, arXiv:astro-ph/13073884
\bibitem[Dorn et al.\ 2014]{Do14}{Dorn, S., Ramirez, E., Kunze, K.E., Hofmann, S., En{\ss}lin, T.A.} 2014, arXiv:astro-ph/14035067
\bibitem[Dorn et al. 2013(2)]{Do13(2)}{Dorn, S., Oppermann, N., En{\ss}lin, T.A.} 2013, \textit{Phys.~Rev.~E}~88~053303, arXiv:astro-ph/13073889


\bibitem[Planck 2013(3)]{Pl13(3)}{Planck Collaboration} 2013, arXiv:astro-ph/13035082 %Inflation
\bibitem[Bartolo et al. 2002]{Ba02}{Bartolo, N., Liddle, A.R.} 2002, \textit{Phys.~Rev.~D}~65~121301, arXiv:astro-ph/0203076
\bibitem[Sasaki et al. 2006]{Sa06}{Sasaki, M., V\"aliviita, J., Wands, D.} 2006, \textit{Phys.~Rev.~D}~74~103003, arXiv:astro-ph/0607627


\end{thebibliography}
\end{document}